\documentclass[fleqn,12pt,twoside]{article}
\usepackage{espcrc1}

\usepackage{graphicx}
\usepackage[figuresright]{rotating}

\newcommand{\ee}{e^+e^-}
\newcommand{\epep}{e^+e^+}
\newcommand{\emem}{e^-e^-}
\newcommand{\eep}{$e^+e^-$-pair}
\newcommand{\eeps}{$e^+e^-$-pairs}
\newcommand{\pipi}{$\pi^+\pi^-$}
\newcommand{\pipips}{$\pi^+\pi^-$-pairs}

\newcommand{\AmS}{{\protect\the\textfont2
  A\kern-.1667em\lower.5ex\hbox{M}\kern-.125emS}}

\title{Latest Results from CERES/NA45}

\author{Johannes P. Wessels\address[gsi]{GSI, Darmstadt, Germany} for the CERES/NA45 Collaboration\\ \vskip8mm
D.~Adamov\'a\address[rez]{NPI ASCR, \v{R}e\v{z}, Czech Republic}, 
G.~Agakichiev\addressmark[gsi],
{H.~Appelsh\"auser}\address[hei]{Heidelberg University, Germany}, 
{V.~Belaga}\address[dub]{JINR Dubna, Russia}, 
{P.~Braun-Munzinger}\addressmark[gsi], 
{R.~Campagnolo}\addressmark[hei],
{A.~Castillo}\addressmark[gsi],
{A.~Cherlin}\address[wis]{Weizmann Institute, Rehovot, Israel}, 
{S.~Damjanovi\'c}\addressmark[hei], 
{T.~Dietel}\addressmark[hei], 
{L.~Dietrich}\addressmark[hei], 
{A.~Drees}\address[sb]{SUNY at Stony Brook, U.S.A.}, 
{S.\,I.~Esumi}\addressmark[hei], 
{K.~Filimonov}\addressmark[hei], 
{K.~Fomenko}\addressmark[dub],
{Z.~Fraenkel}\addressmark[wis], 
{C.~Garabatos}\addressmark[gsi], 
{P.~Gl\"assel}\addressmark[hei], 
{G.~Hering}\addressmark[gsi], 
{J.~Holeczek}\addressmark[gsi], 
{V.~Kushpil}\addressmark[rez], 
{B.~Lenkeit}\address[cern]{CERN, Geneva, Switzerland}, 
{W.~Ludolphs}\addressmark[hei],
{A.~Maas}\addressmark[gsi], 
{A.~Mar\'{\i}n}\addressmark[gsi], 
{F.~Messer}\addressmark[sb],
{J.~Milo\v{s}evi\'c}\addressmark[hei],
{A.~Milov}\addressmark[wis], 
{D.~Mi\'skowiec}\addressmark[gsi], 
{L.~Musa}\addressmark[cern],
{Yu.~Panebrattsev}\addressmark[dub], 
{O.~Petchenova}\addressmark[dub], 
{V.~Petr\'a\v{c}ek}\addressmark[hei], 
{A.~Pfeiffer}\addressmark[cern], 
{J.~Rak}\address[mpi]{MPI, Heidelberg, Germany}, 
{I.~Ravinovich}\addressmark[wis], 
{P.~Rehak}\address[bnl]{BNL, Upton, U.S.A.}, 
{M.~Richter}\addressmark[hei], 
{H.~Sako}\addressmark[gsi], 
{W.~Schmitz}\addressmark[hei], 
{J.~Schukraft}\addressmark[cern], 
{S.~Sedykh}\addressmark[gsi], 
{W.~Seipp}\addressmark[hei], 
{A.~Sharma}\addressmark[gsi], 
{S.~Shimansky}\addressmark[dub], 
{J.~Sl\'{\i}vov\'a}\addressmark[hei],
{H.\,J.~Specht}\addressmark[hei], 
{J.~Stachel}\addressmark[hei], 
{M.~\v{S}umbera}\addressmark[rez], 
{H.~Tilsner}\addressmark[hei], 
{I.~Tserruya}\addressmark[wis], 
{J.\,P.~Wessels}\addressmark[gsi], 
{T.~Wienold}\addressmark[hei], 
{B.~Windelband}\addressmark[hei], 
{J.\,P.~Wurm}\addressmark[mpi], 
{W.~Xie}\addressmark[wis], 
{S.~Yurevich}\addressmark[hei], 
{V.~Yurevich}\addressmark[dub]}

\begin{document}

\maketitle

\begin{abstract}
In this talk latest results from the analysis of \eeps~emitted in Pb$+$Au
collisions at 40~AGeV/c and a combined analysis of all data
available at 158~AGeV/c are presented. The 
enhancement of low-mass \eeps~($m_{ee}>$0.2~GeV/c$^2$) 
with respect to the expected yield from hadron decays
first reported at 158~AGeV/c is also found at 40~AGeV/c and is even 
larger there. 
Comparing to various theoretical models based on \pipi~annihilation,
the data can only be reproduced, if the properties of the intermediate $\rho$ 
in the hot and dense medium are modified.
Theoretically, the modification
is linked to baryon density rather than temperature. Constraints from 
hadron data taken at the same beam energies indeed 
indicate a fireball evolution along
a trajectory of higher baryon density at 40 AGeV/c, consistent with the 
observed larger enhancement factor.
\end{abstract}

\section{INTRODUCTION}

Along with the deconfinement phase transition to the quark gluon 
plasma (QGP), chiral symmetry is expected to be restored. 
The energy density at which the phase transition is predicted by two-flavor 
lattice calculations is $\epsilon \approx 0.7$~GeV/fm$^3$~\cite{karsch} with
large errors due to uncertainties in the critical temperature.
Simple estimates employing the Bjorken expansion scenario~\cite{bjorken}
show that the initial energy density reached in central Pb$+$Pb collisions 
is about 3~GeV/fm$^3$ at the full SPS energy. The totality of the observables 
has led to the announcement that a new state of matter is indeed formed in 
central heavy ion collisions at the SPS~\cite{cern}.
The measurement of \eeps~ is particularly well suited to study this matter
since they are emitted throughout the entire collision and are 
themselves not subject to the strong final state interaction.
In this context, the decay of the $\rho$ vector meson into \eeps~is of 
particular interest, due to its direct
link to chiral symmetry restoration. 
Because of the short lifetime (1.3~fm/c), \eeps~ from $\rho$ decays are 
dominated by decays within the hot and dense medium, directly
mirroring its properties.
Experimentally, however, the study of \eeps~
is notoriously difficult, because of the small production cross
section as well as the large combinatorial background in heavy ion
collisions.

In the following we briefly describe the experimental setup, followed by 
a description of the electron analysis procedure. We then present 
results from a combined analysis of all the data that were taken at the
full SPS energy of 158~AGeV. Then the final results of the data taken
at 40~AGeV are shown. We conclude after a comparison of the data 
to models and a discussion of the fireball evolution. 

\section{CERES SETUP}

The CERES spectrometer is optimized to measure \eeps~in the mass range
up to about 1~GeV/c$^2$ in the range 2.1$<\eta<$2.65 close to 
midrapidity. A cross section through the azimuthally symmetric setup
is shown in Figure~1.

\begin{figure}[htb]
\includegraphics[angle=-90,width=\textwidth, clip=]{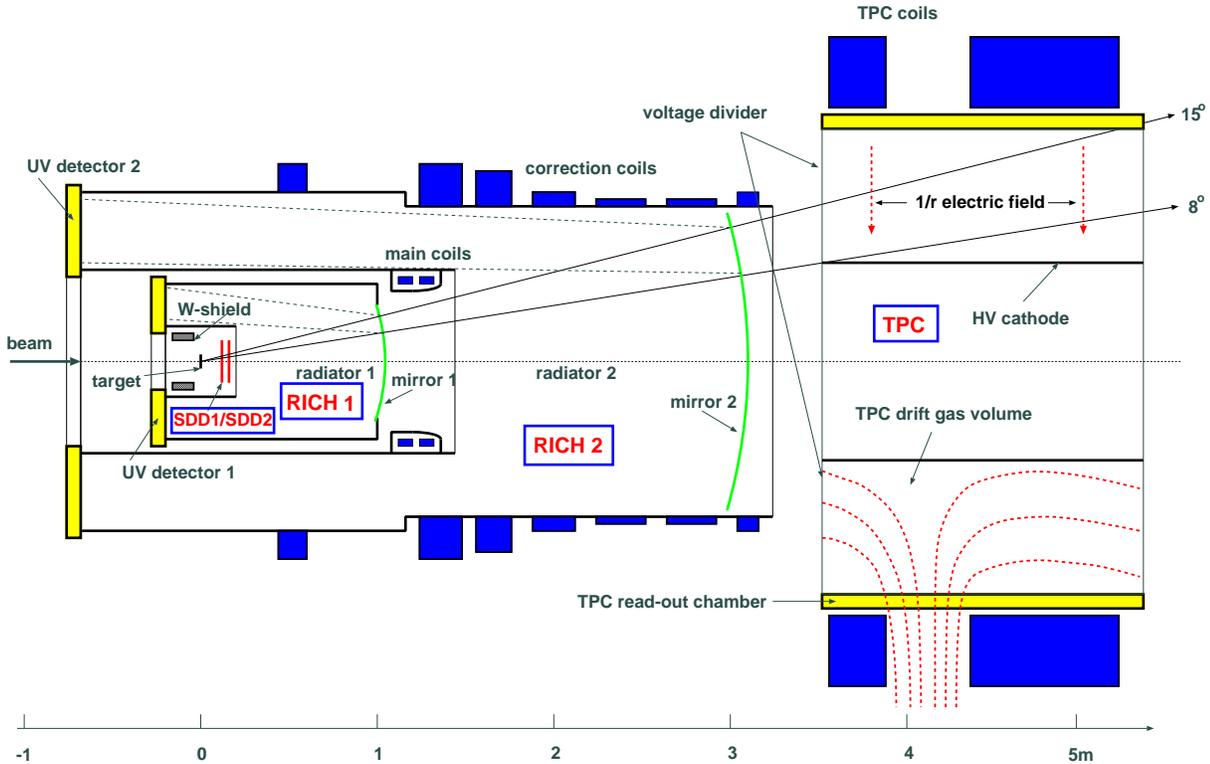}
\label{setup}
\vspace*{-23pt}
\caption{Cross section through the upgraded setup of the CERES spectrometer}
\end{figure}

A segmented micro target (8 disks of Au of 25~$\mu$m thickness and
600~$\mu$m diameter evenly spaced along 22~mm)
is followed by a set of two {\bf S}ilicon {\bf D}rift {\bf D}etectors (SDD1/SDD2) at
a distance of 10~cm and 13~cm respectively. The segmentation minimizes conversions
of photons in the target. The set of silicon drift 
detectors provides precise determination of the angles of all charged
particles, determination of the event vertex, and the total charged
particle multiplicity for each event. Along with the position measurement
the specific energy loss is recorded for each charged particle.

Electrons are identified with two {\bf R}ing {\bf I}maging {\bf Ch}erenkov
counters (RICH1/2) which are operated at a threshold of $\gamma_{th}=32$
suppressing more than 99\% of all charged hadrons. In 1995/96 the spectrometer
was operated with a radial magnetic field provided by two air coils
with counter-running current
in between the two RICHes~\cite{ceres-setup}. With this setup a mass resolution 
of about 5.7\% was reached in the mass region of the $\rho/\omega$.

For the runs in 1999/2000 the experiment was upgraded by the addition of 
a radial TPC~\cite{amqm00} inside a separate radial magnetic field
(B$_{\rm max}$=0.5~T). Over the
total length of 2~m a maximum of 20 hits are recorded for each charged 
particle track. In the 1999 analysis 
a mass resolution of about 4.2\% was achieved
in the $\rho/\omega$ region. For the 2000 data the goal is to reach 
with better calibration about
2\%. For runs with the TPC the magnetic field between the RICHes was
switched off and electrons were identified by combining the two RICHes,
improving the electron efficiency to about 90\% (as compared to about 
70\% in 1995/1996).

A compilation of all Pb$+$Au events taken with different
configurations of the CERES spectrometer is shown in Table~\ref{data}.

\begin{table}[htb]
\caption{Number of collected Pb$+$Au events along with the respective trigger cross sections.}
\label{data}
\setlength{\tabcolsep}{0.5cm}
\begin{tabular}[t]{rlcl}
Year & $p_{beam}$&  $\sigma/\sigma_{geo}$ & Events \\ \hline \\
1995 &  158~AGeV/c &   35\% & 8.5$\cdot 10^6$ \\
1996 &  158~AGeV/c &   30\% & 43$\cdot 10^6$ \\ \\
1999 &   40~AGeV/c &   30\%& 8$\cdot 10^6$ - partial readout\\ \\ 
2000 &   80~AGeV/c &  $<$30\%& 0.5$\cdot 10^6$ \\
     &  158~AGeV/c &   8(25)\% & 33(3.4)$\cdot 10^6$ \\
\end{tabular}\\[2pt]
\end{table}

The data taken in 1995 and 1996 used the original CERES setup without the
TPC. All data taken thereafter employed the upgraded setup shown in
Figure~1. The new read-out of the experiment
was only partially operational for the low energy run at
40~AGeV/c in 1999 limiting both statistics and performance. 
The large data sample taken in 2000 at the full 
SPS energy is still being analyzed.

\section{ELECTRON ANALYSIS}
\label{ana}

\begin{figure}[h]
  \begin{minipage}[t]{0.48\textwidth}
    \vspace{8pt}
    \centering
    \includegraphics[height=65mm,clip=]{./figs/leer.eps}%
    \vspace*{-7pt}
        \vspace*{-23pt}         
    \caption{\label{dedxtpc} Specific energy loss in the TPC for electron candidates
        identified with the RICHes. The indicated cut was used to efficiently
        reject a remaining contamination from pion tracks. 
        The cut at p=1~GeV/c corresponds to the $p_t$-cut of 0.2~GeV/c.}
  \end{minipage}%
  \hfill
  \begin{minipage}[t]{0.48\textwidth}
    \vspace{0pt}
    \centering
    \includegraphics[height=80mm, angle=-90, clip=]{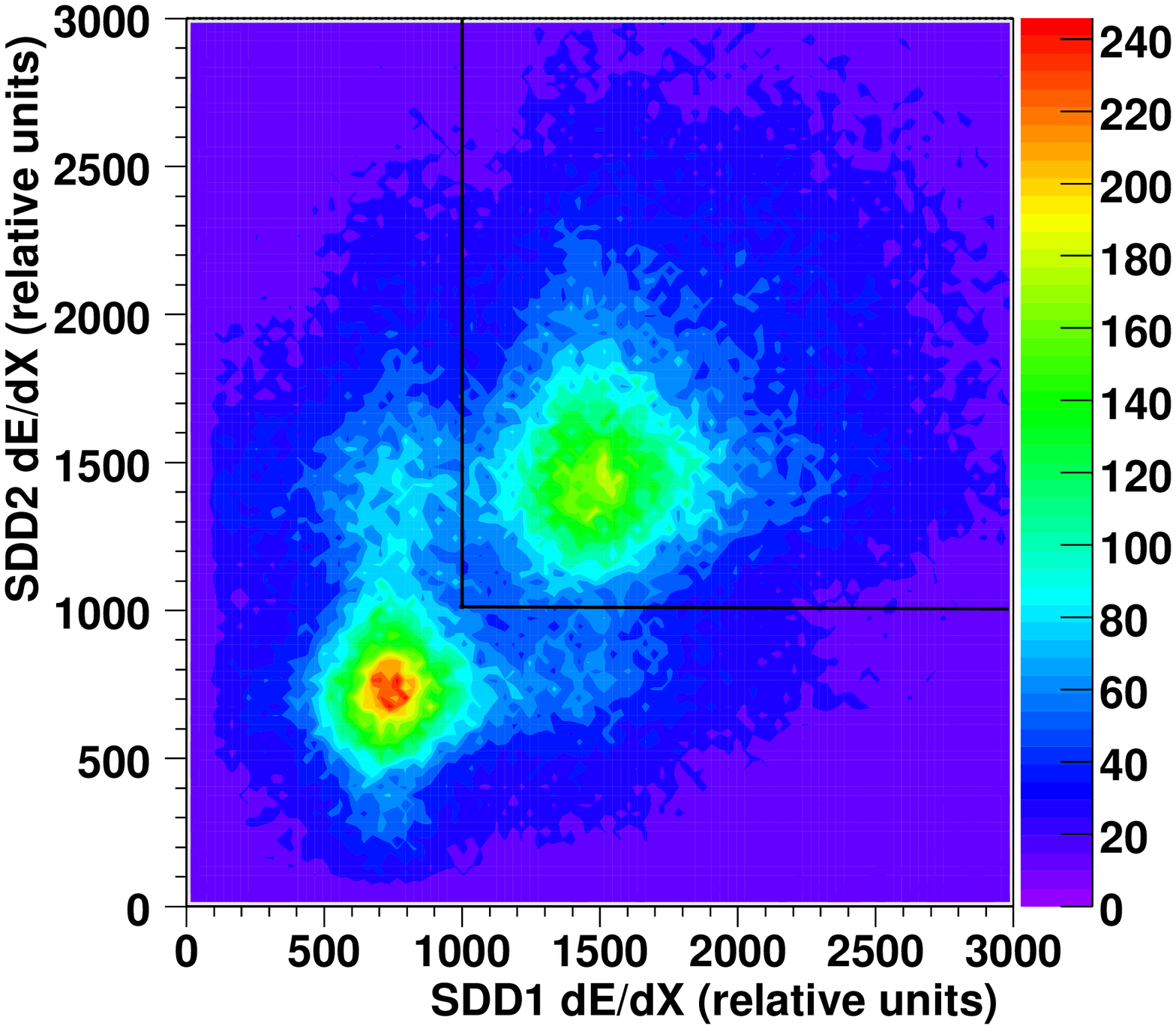}%
        \vspace*{-23pt}
    \caption{\label{dedxsdd} Specific energy loss for charged particle tracks
in the two silicon drift detectors. The cut indicates unresolved pairs of 
tracks mostly due to conversions.}
  \end{minipage}
\end{figure}

While \eeps~are an attractive probe, they are notoriously hard to measure
due to the small production cross section and the enormous combinatorial
background from unrecognized photon conversions and $\pi^0$ Dalitz decays. 
The analysis steps for the 1995/96 data are detailed in Ref.~\cite{combined9596}.
Here, we only describe the analysis of data taken with the TPC and the field
between the RICHes turned off. Electron candidates are
selected when ten or more photon hits in RICH1 and RICH2 form a ring with
asymptotic radius. To those candidates a transverse momentum cut of 200
MeV/c is applied reducing contributions from $\pi^0$ Dalitz decays and
conversions by about a factor of 10. Electron identification is further
improved,removing high-$p_t$ pions
and accidental matches between the RICHes and the TPC,
by a cut on specific energy loss in the TPC vs. momentum as
shown in  Figure~\ref{dedxtpc}.
Those conversions and Dalitz pairs with very small opening angles 
that do not lead to separate rings in the RICHes are efficiently removed by 
rejecting tracks with large specific energy loss in both silicon 
drift detectors as shown in Figure~\ref{dedxsdd}. Further, electron candidates
without a match in the TPC within 70~mrad of an electron track are also removed.
The complete set of cuts employed in the analysis is shown in detail in
Ref.~\cite{qm01,pre40,sd}. 
Finally, the invariant mass spectrum of \eeps~is obtained by subtracting 
twice the geometric average of the remaining
like-sign pairs from the remaining unlike-sign pairs 
($N_{\ee}-2\sqrt{N_{\epep}\cdot N_{\emem}}$).

\section{Pb$+$Au AT 158~AGeV/c}

\subsection{Invariant mass spectrum of \eeps}
\label{mass158}

Analyses of the invariant mass spectra of \eeps~from the runs in 1995/1996
have been published~\cite{95data,qm97,qm99}. In all cases cuts on the
pair opening angle ($\Theta_{ee}>35$~mrad) and the transverse momentum of
the single electron track ($p_t>200$~MeV/c$^2$) have been applied, and
the spectra were compatible within their respective statistical and systematic
errors.
Recently \cite{combined9596}
the data of 1995 and 1996 were combined in a unified analysis approach. 
For the two runs the trigger cross-section was slightly different. 
The cross section for the combined data corresponds to the 32\% most
central events. The average charged particle multiplicity  
is $\langle N_{ch}\rangle=\langle dN_{ch}/d\eta\rangle
\cdot \Delta\eta = 245\cdot 0.55=135$ for the quoted rapidity interval.
Since the efficiency for detecting \eeps~ depends on
multiplicity, an efficiency correction is applied on an event-by-event
basis. In the combination, the data were averaged bin-by-bin  
with weights according to the inverse squares of the relative errors of 
the individual measurements.

\begin{figure}[htb]
  \begin{minipage}[t]{0.48\textwidth}
    \vspace{0pt}
    \centering
    \includegraphics[width=\textwidth, angle=0, clip=]{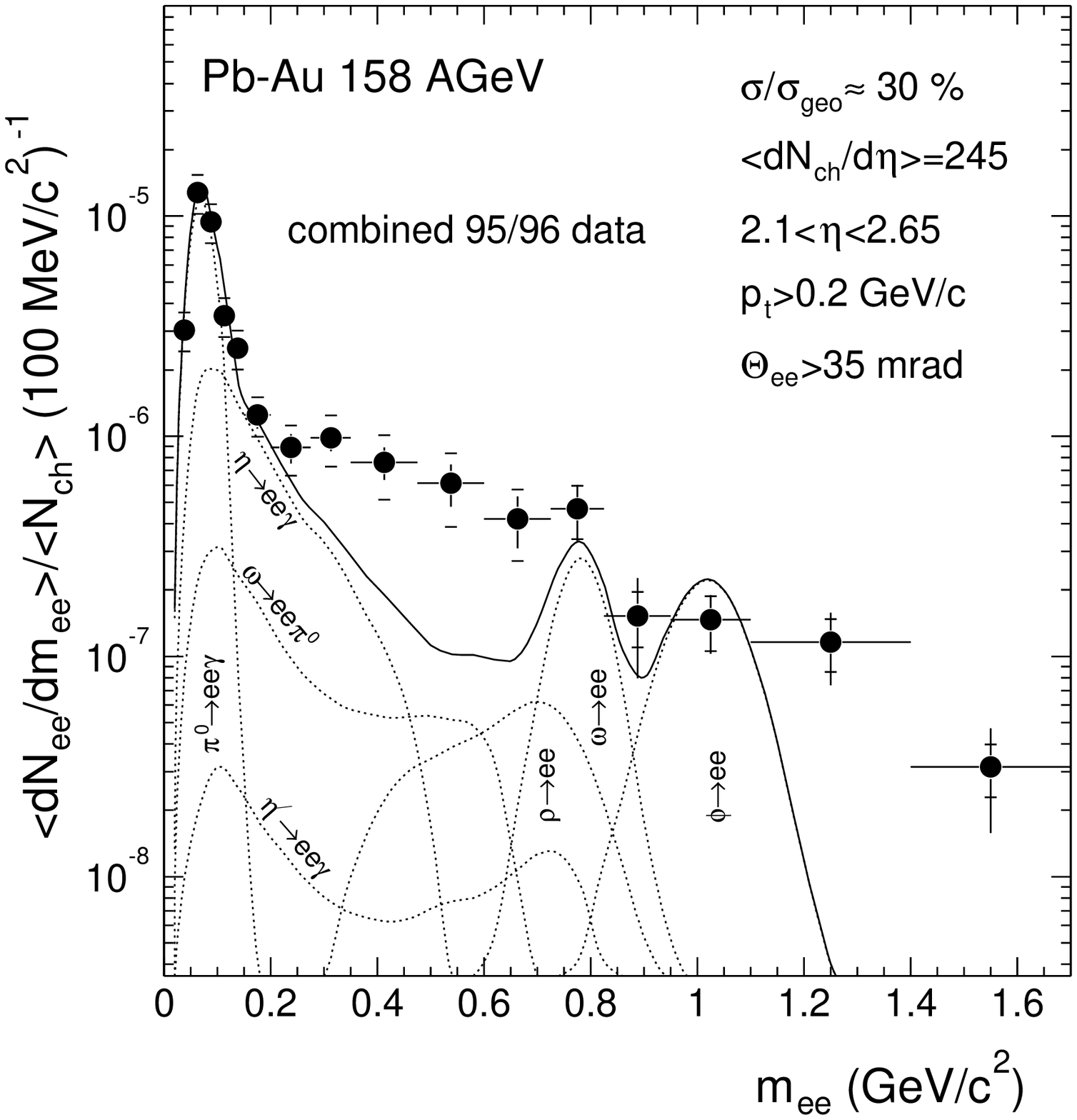}%
        \vspace*{-23pt}
    \caption{\label{mass9596} Invariant mass spectrum of \eep~emitted in 158~AGeV/c 
      Pb$+$Au collisions from the combined analysis of 1995 and 1996 data. The solid 
      line shows the expected yield from hadron decays, dashed lines indicate the
        individual contributions to the total yield. See text for a discussion of
        systematic errors.}
  \end{minipage}%
  \hfill
  \begin{minipage}[t]{0.48\textwidth}
    \vspace{0pt}
    \centering
    \includegraphics[width=\textwidth,clip=]{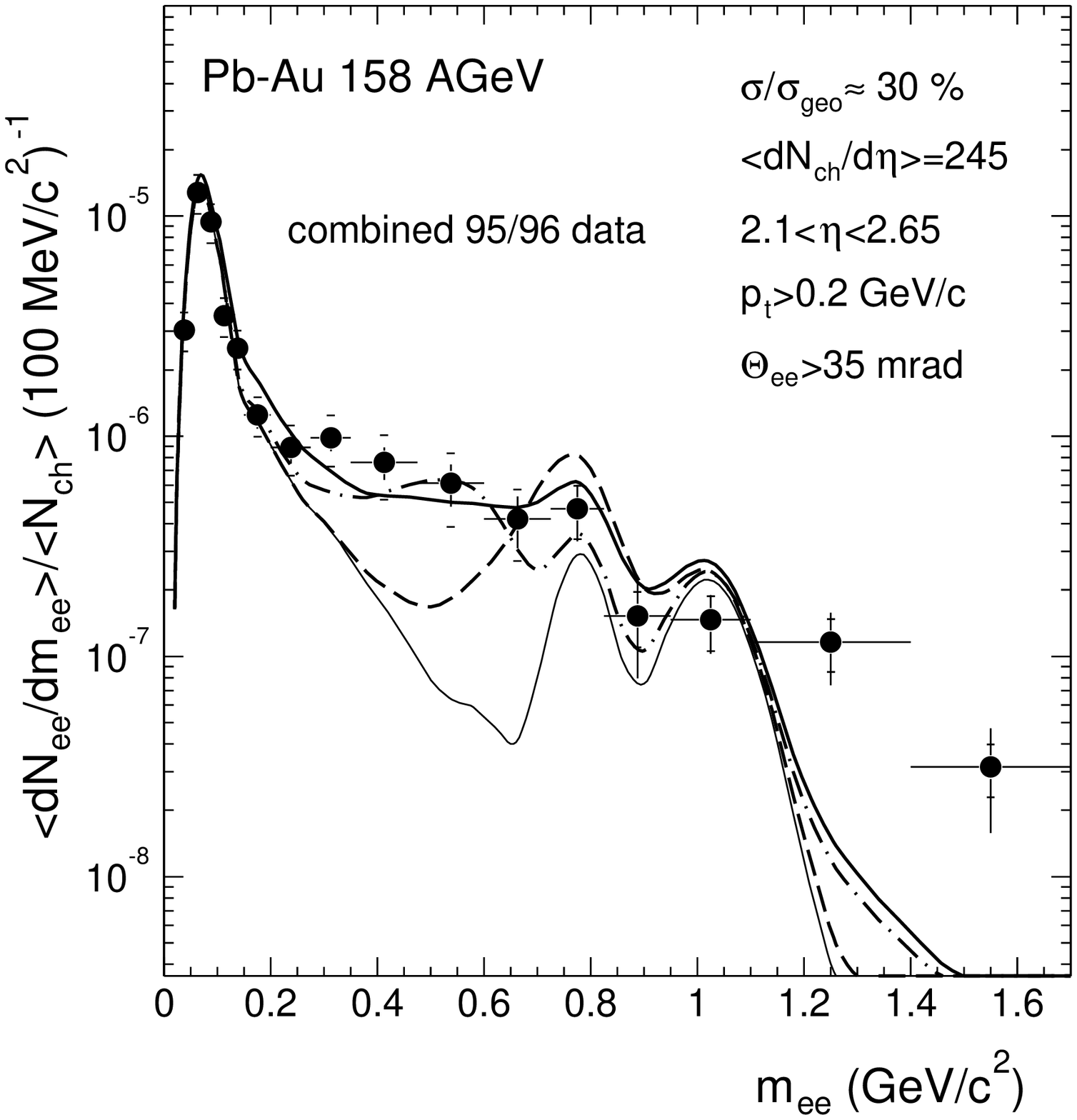}%
        \vspace*{-23pt}
    \caption{\label{mass9596t} Comparison of the experimental data 
      to i) free hadron decays without $\rho$ decays (thin solid line), 
      ii) model calculations 
      with a vacuum $\rho$ spectral function (thick dashed line), 
      iii) with dropping in-medium $\rho$-mass (thick dash-dotted line, iv)
      with a medium-modified $\rho$ spectral function (thick solid line).}
  \end{minipage}
\end{figure}

The combined data are shown in Figure~\ref{mass9596}. The yield of pairs 
with $m_{ee}>0.2$~GeV/c$^2$ for the entire data sample is 2666$\pm$260
with a signal-to-background ratio of 1/12.
On top of the statistical errors, the spectrum has 
systematic errors of 20\% for $m_{ee}<0.2$~GeV/c$^2$ and 28\% on average for 
$m_{ee}>0.2$~GeV/c$^2$, larger here
due to the background subtraction and smoothing procedure.
The background smoothing procedure introduces an additional 7\%
uncertainty in the overall normalization of the spectrum for
m$_{ee}>$0.2~GeV/c.
A detailed description of all errors can be found in Ref.~\cite{sd,combined9596}

In Figure~\ref{mass9596} the data are compared to the yield from
hadron decays obtained from a {\em cocktail} containing particle ratios
from a statistical model fitted to measured particle ratios in Pb$+$Pb
collisions. Rapidity and transverse momentum
distributions of the particles are taken from 
systematics of Pb$+$Pb collisions. The resulting distribution has been 
folded with the experimental mass resolution (5.7\% in the $\omega$ region). 
In the region $m_{ee}>0.2$~GeV/c$^2$
the data exceed the yield from the {\em cocktail} by a factor of
2.3$\pm$0.2(stat)$\pm$0.6(syst). An additional systematic error
of 0.7 is connected to the decay {\em cocktail}. The enhancement 
is not observed in proton induced reactions~\cite{pp}.

An independent analysis was performed using event mixing to
generate the background~\cite{gh}.
The shape of this background is found to be identical
to that of like-sign pairs, and the resulting invariant mass spectrum 
agrees well with the one in Figure~\ref{mass9596}. In addition,
a lower single electron $p_t$-cut $>$ 100~MeV/c$^2$ 
was used resulting in
a factor of two more statistics for \eep~ masses above 0.2~GeV/c$^2$ at a
factor of three worse signal-to-background ratio. 
Again the resulting invariant mass spectrum agrees well with the one shown  
in Figure~\ref{mass9596}.

In Figure~\ref{mass9596t} the data are compared to the yield from
various models based on \pipi~annihilation with an intermediate $\rho$ 
vector meson. The hadron decay {\em cocktail} 
is added without any contribution from the $\rho$ to avoid double counting (thin line).
Using the vacuum $\rho$ spectral function produces the
thick dashed line. The other two model calculations both contain
modifications of the $\rho$ spectral function. They involve in-medium 
$\rho$ spectral functions 
with either a dropping $\rho$ mass~\cite{br} (thick dash-dotted line) or with a 
spread $\rho$ width~\cite{rw} (thick line). Clearly, 
a modification of the $\rho$ properties in the medium is needed to reach
agreement with the experimental data. Unfortunately,
the data do not allow to discern between the two different scenarios.

\begin{figure}[htb]
  \begin{minipage}[t]{0.72\textwidth}
    \vspace{0pt}
    \centering
    \includegraphics[width=\textwidth, angle=0,clip=]{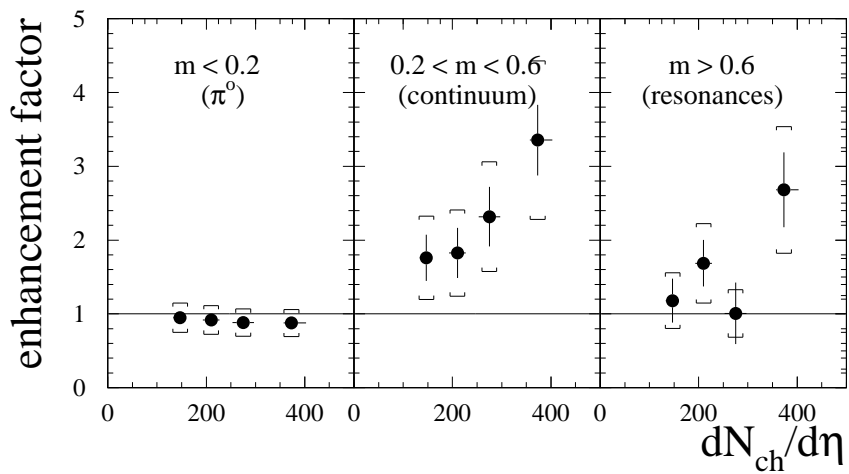}%
  \end{minipage}%
  \hfill
  \begin{minipage}[t]{0.26\textwidth}
    \vspace{-20pt}
    \caption{\label{mult} Enhancement of \eeps ~from Pb$+$ Au collisions at 158~AGeV/c
      with respect to the yield expected from the hadron decay {\em cocktail}
      as a function of multiplicity and pair mass.}
  \end{minipage}
\end{figure}

\subsection{Multiplicity dependence of the enhancement}

The combined data sample was subdivided into four different multiplicity
bins. For each of these bins and
for three different intervals in invariant mass, the 
ratio of the integral yield of \eeps~to the expected yield from the
hadron decay {\em cocktail} was evaluated. The results are shown in 
Figure~\ref{mult}. The mass region below 0.2~GeV/c$^2$, dominated 
by $\pi^0$ Dalitz decays, is well reproduced by the {\em cocktail} 
for all centralities. In the mass region  0.2~GeV/c$^2<m_{ee}<$0.6~GeV/c$^2$
the enhancement with respect to the {\em cocktail} rises 
approximately linearly with 
multiplicity indicating that the rise of the integrated \eep~yield 
is consistent with a quadratic multiplicity dependence;
this is roughly expected for a two-body process like \pipi~annihilation.

\section{Pb$+$Au AT 40~AGeV/c}

In 1999 the SPS delivered a 40~AGeV/c Pb-beam when the upgraded CERES spectrometer
had just been commissioned. Preliminary results of what is presented in the
following have already previously been shown \cite{qm01,pre40}. The data contain 
central events with an average multiplicity density at this energy
of $\langle dN_{ch}/d\eta\rangle=216$.

\subsection{Invariant mass spectrum of \eeps}

\begin{figure}[htb]
  \begin{minipage}[t]{0.65\textwidth}
    \vspace{0pt}
    \centering
    \includegraphics[width=\textwidth, angle=0,clip=]{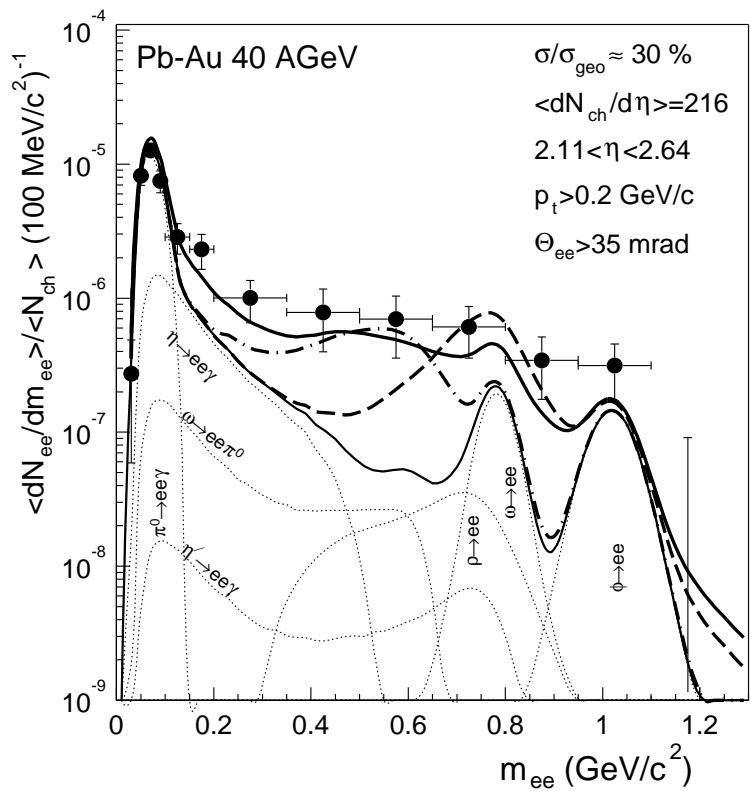}%
  \end{minipage}%
  \hfill
  \begin{minipage}[t]{0.33\textwidth}
    \vspace{-20pt}
    \caption{\label{mass40} \eep~mass spectrum from Pb$+$Au collisions at 40~AGeV/c
      compared to free hadron decays and various model calculations 
        (line codes as in Fig.~\ref{mass9596t}). Errors shown are only statistical,
        see text for a discussion of systematic errors.
      }
  \end{minipage}
\end{figure}

The analysis leading to the invariant mass spectrum for the 40~AGeV/c data
shown in Figure~\ref{mass40} has been outlined in Section~\ref{ana} and is 
described in detail in~\cite{sd}. 
The resulting spectrum contains 249$\pm$28 \eeps~ in the mass range 
below 0.2~GeV/c$^2$ at a signal-to-background ratio of 1/1. The integrated
yield for high mass pairs above 0.2~GeV/c$^2$ is 185$\pm$48 pairs at
a signal-to-background ratio of 1/6. The systematic errors are 16\% and
20\% for the low and high mass region, respectively. 

For the comparison to the hadron decay {\em cocktail} the particle ratios
have again been taken from statistical model fits adjusted to measured ratios 
from Pb$+$Pb collisions at 40~AGeV/c. Also
the rapidity distributions and transverse momentum spectra follow the
systematics of measurements at 40~AGeV/c. Again the cocktail has been folded
with the experimental mass resolution (4.2\% in the $\omega$ region).

In the low mass region $m_{ee}<0.2$~GeV/c$^2$ excellent
agreement between the data and the {\em cocktail} is obtained. Here the
ratio of data/decays is 0.98$\pm$0.11(stat)$\pm$0.16(syst).
In the high mass region $m_{ee}>0.2$~GeV/c$^2$, 
however, the data exceeds the {\em cocktail} by a 
factor of 5.1$\pm$1.3(stat)$\pm$1.0(syst) with an additional 
systematic error of 1.5 from the {\em cocktail}.
This enhancement is larger than that observed at 
full SPS energy as quoted above, while the yield
itself is not much larger than at 158~AGeV/c. 
Note that in the comparison of the enhancement factors 
between the two energies the additional systematic uncertainty
from the {\em cocktail} essentially drops out.

Along with the data, model {\em predictions} \cite{rapp02} are presented 
for this energy. Again, a treatment of the $\rho$ unmodified
by medium effects is clearly ruled out. However, once more the data do not
allow to discern the two scenarios involving in-medium modifications 
of the $\rho$ propagator with a shifted mass, i.e. 
{\em Brown/Rho scaling}~\cite{br,rapp02},
or with a spread width~\cite{rapp02}.

\begin{figure}[htb]
  \begin{minipage}[t]{0.48\textwidth}
    \vspace{0pt}
    \centering
    \includegraphics[width=0.95\textwidth, angle=0, clip=]{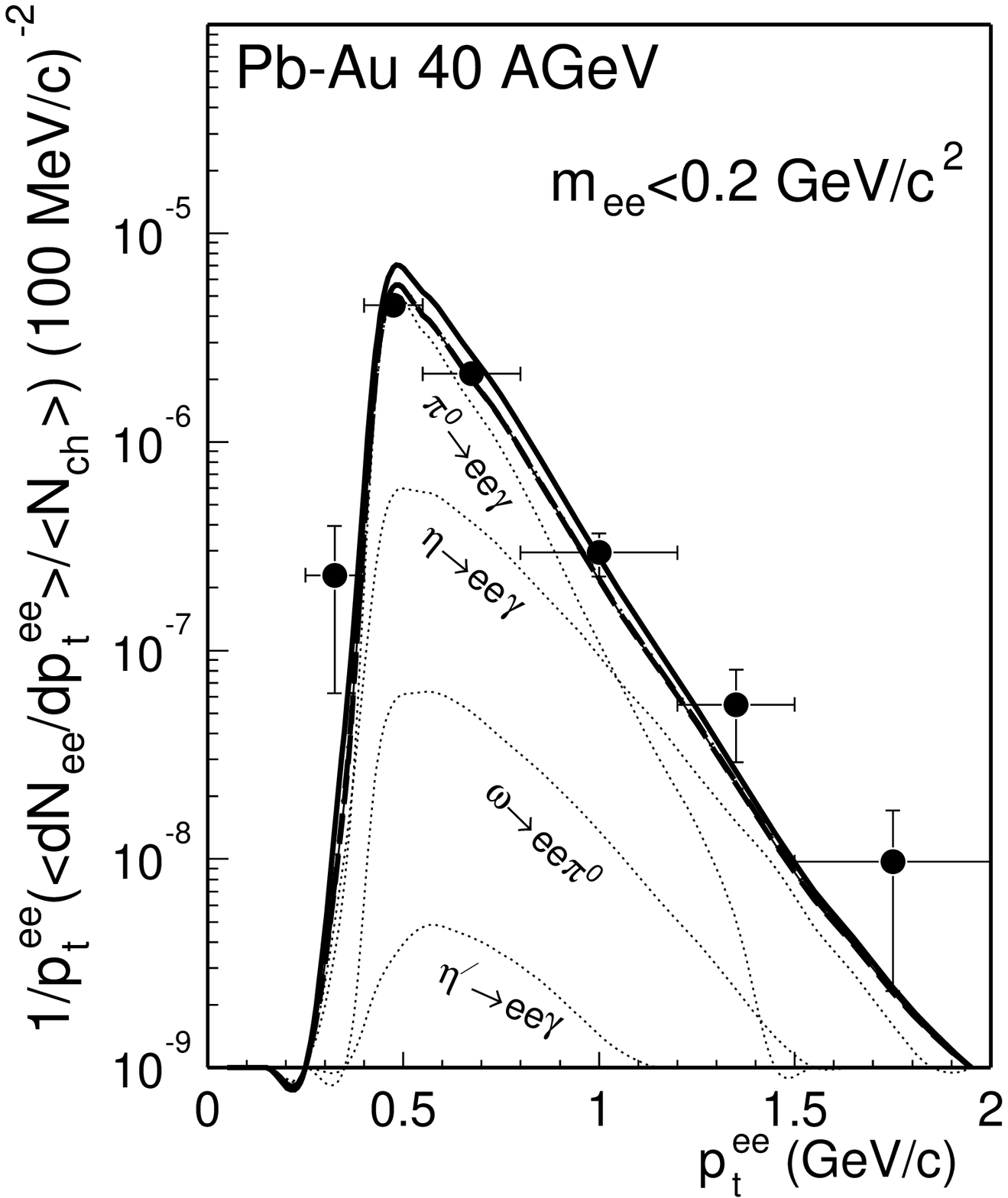}%
        \vspace*{-23pt}
  \end{minipage}%
  \hfill
  \begin{minipage}[t]{0.48\textwidth}
    \vspace{0pt}
    \centering
    \includegraphics[height=88mm,clip=]{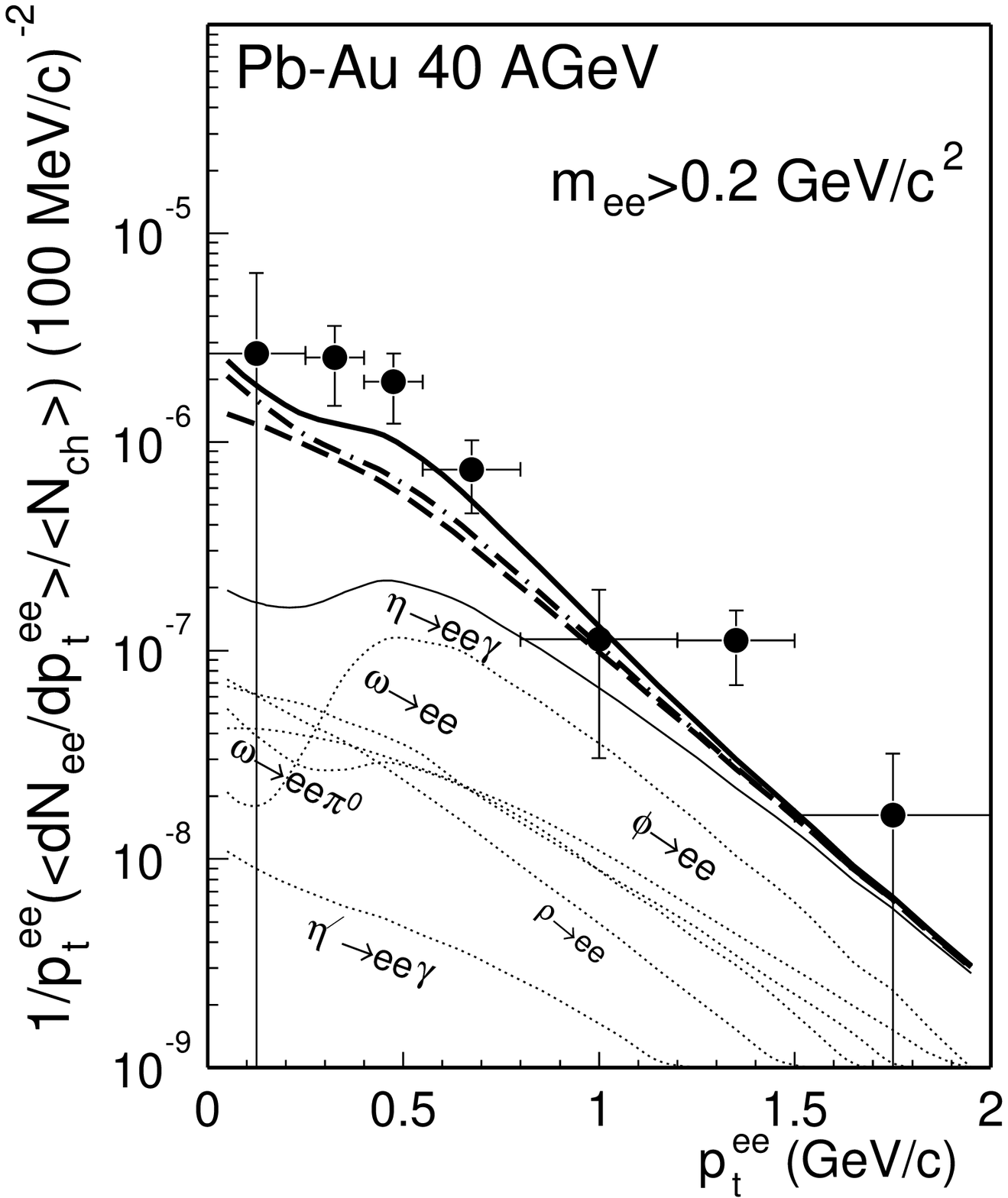}%
        \vspace*{-23pt}
  \end{minipage}
    \caption{\label{pt40} Invariant pair-$p_{t}^{ee}$ spectra for \eep~masses 
      m$_{ee}$$<$0.2 GeV/c$^{2}$ (left panel) and m$_{ee}$$>$0.2 GeV/c$^{2}$ 
        (right panel) compared to hadron decays and model calculations
      (line codes and errors as in Fig.~\ref{mass9596t}).}
\end{figure}

\subsection{Invariant transverse momentum spectra of \eeps}

The invariant differential pair tranverse momentum spectra ($p_t^{ee}$)
are shown in Figure~\ref{pt40} for masses 
$m_{ee}<$0.2~GeV/c$^2$ and $m_{ee}>$0.2~GeV/c$^2$ respectively.
It can be seen that the Dalitz region with $m_{ee}<$0.2~GeV/c$^2$
is well reproduced by the hadron decay {\em cocktail} without significant 
contributions from pion annihilation. 
The steep cut-off below $p_t^{ee}<$0.4~GeV/c is due to the minimum
transverse momentum cut of 0.2~GeV/c for the single electron. 
The pair transverse momentum distribution for pairs with 
$m_{ee}>$0.2~GeV/c$^2$, however, is strongly enhanced for pairs with 
$p_t^{ee}<$0.6~GeV/c compared to the {\em cocktail}.
Here, all model calculations (also the one using the
vacuum $\rho$ spectral function) describe the pair transverse momentum
spectra reasonably well. Therefore, the shape of the spectrum appears
to mostly exhibit the basic feature of \pipi~annihilation 
rather than medium effects, while the latter affect the shape of the invariant 
mass spectrum.  

\section{DISCUSSION OF RESULTS}

We find, both at 158~AGeV/c as well as at 40~AGeV/c, that the production 
of \eeps~ with $m_{ee}>$0.2~GeV/c$^2$ is significantly enhanced
as compared to expectations from hadron decays. 
Moreover, the enhancement is larger
at 40~AGeV/c compared to 158~AGeV/c. In both cases, theories based
on \pipi~annihilation are only able to describe the data, if 
medium modifications of the $\rho$ propagator are included 
\cite{br,rw,rw2,rapp02}. 

Comparison of the two energies sheds light on the
interesting question whether changes in the properties of the
$\rho$ propagator are more sensitive to changes in the temperature
or changes of the baryon density as favored by theory~\cite{br,rw,rw2,rapp02}. 
In order to evaluate the e$^+$e$^-$-rates one would
need complete knowledge of the  dynamical
evolution of the fireball. There is no direct handle on the 
time spent in the various stages of the collision (with the exception
of the emission duration of at most 2-3~fm/c and the thermal freeze-out time of
6-8~fm/c from HBT measurements~\cite{qm02hbt,hbtpaper}).
The evolution in terms of temperature and baryochemical
potential can, however, be determined at chemical and thermal freeze-out.
At chemical freeze-out the total baryon density can be estimated
from statistical model calculations fitted to measured particle
ratios. It is found to be approximately 0.75 in units of normal nuclear 
matter density $\rho_0$ at associated temperatures of 168(145)~MeV for 
158(40)~AGeV/c~\cite{js}.
The baryon density at thermal freeze-out can be derived from the 
measured HBT-radii~\cite{qm02hbt,hbtletter,hbtpaper} and
rapidity densities~\cite{na44}. It is 0.34(0.53) at 
temperatures of 120(120)~MeV for 158(40)~AGeV/c.
Note that after chemical freeze-out these baryon densities are all below $\rho_0$.
Adding in results on initial baryon densities from transport 
calculations~\cite{bengt}, the baryon density appears to be 
consistently higher all the way along the fireball trajectory at the lower
beam energy, including a part before reaching chemical equilibrium.

Theoretical calculations which evolve the system through such
higher baryon densities
are in agreement with the observed increased enhancement at the lower beam energy
\cite{rapp02}.
Taken together, these observations therefore
support the importance of coupling the 
$\rho$ propagator in the medium to baryon density 
rather than temperature.

The TPC calibration for the 2000 data has been completed; 
the analysis of the large number of events taken at the full SPS energy is about
to start. With this data it should also be possible to study the
behavior of the $\omega$ and the $\phi$. It would further be
interesting to compare the invariant mass spectrum of \pipips~\cite{facchini} 
to that from \eeps, which may allow to discriminate between 
earlier and later stages of the collision.

\section{ACKNOWLEDGMENTS}

We are grateful for support by the German BMBF, the U.S. DoE, the Israeli 
Science Foundation, and the MINERVA Foundation.

\end{document}